\documentclass[prl,twocolumn,floatfix]{revtex4}
\usepackage{graphicx}

\newlength{\onecolfig}
\newlength{\twocolfig}
\newcommand{\ion}[2]{\mbox{$^{#2}$#1$^+$}}
\newcommand{\Ca}[1]{\ion{Ca}{#1}}

\newcommand{\lev}[2]{\mbox{#1$_{\mbox{\tiny$#2$}}$}}
\newcommand{\hfslev}[3]{\mbox{#1$^{\mbox{\tiny$#3$}}_{\mbox{\tiny$#2$}}$}}

\newcommand{\unit}[1]{\,\mbox{#1}}

\newcommand{\MHz}{\unit{MHz}}
\newcommand{\GHz}{\unit{GHz}}

\newcommand{\uW}{\unit{$\mu$W}}

\newcommand{\nm}{\unit{nm}}

\newcommand{\cps}{\unit{s$^{-1}$}}
\newcommand{\persec}{\unit{s$^{-1}$}}
\newcommand{\ms}{\unit{ms}}
\newcommand{\us}{\unit{$\mu$s}}
\newcommand{\ns}{\unit{ns}}

\newcommand{\degree}{\mbox{$^{\circ}$}}
\newcommand{\degC}{\mbox{\degree{}C}}

\newcommand{\etal}{{\em et al.}}

\newcommand{\ish}{\mbox{$\sim$}\,}

\newcommand{\gtish}{\protect\raisebox{-0.4ex}{$\,\stackrel{>}{\scriptstyle\sim}\,$}}

\newcommand{\ket}[1]{\mbox{$\left| #1 \right>$}}

\newcommand{\sub}[1]{\mbox{$_{\mbox{\tiny #1}}$}}

\newcommand{\smalltriangle}{\protect\raisebox{0.2ex}{\tiny $\triangle$}}

\begin{document}
\bibliographystyle{apsrev}

\title{High-fidelity readout of trapped-ion qubits}

\author{A. Myerson, D. Szwer, S. Webster, D. Allcock, M. Curtis, G. Imreh, J. Sherman, D. Stacey, A. Steane and D. Lucas}
\affiliation{Department of Physics, University of Oxford, Clarendon Laboratory, Parks Road, Oxford OX1 3PU, U.K.}

\date{22 April 2008; submitted 12 February 2008}

\begin{abstract}
We demonstrate single-shot qubit readout with fidelity sufficient for fault-tolerant quantum computation, for two types of qubit stored in single 
trapped calcium ions. For an optical qubit stored in the (\lev{4S}{1/2}, \lev{3D}{5/2}) levels of \Ca{40} we achieve 99.991(1)\% average readout 
fidelity in one million trials, using time-resolved photon counting. An adaptive measurement technique allows $99.99\%$ fidelity to be reached in 
145\us\ average detection time. For a hyperfine qubit stored in the long-lived \lev{4S}{1/2} ($F=3$, $F=4$) sub-levels of \Ca{43} we propose and 
implement a simple and robust optical pumping scheme to transfer the hyperfine qubit to the optical qubit, capable of a theoretical fidelity 99.95\% 
in 10\us. Experimentally we achieve 99.77(3)\% net readout fidelity, inferring at least 99.87(4)\% fidelity for the transfer operation.
\end{abstract}


\maketitle

A quantum computer (QC) requires qubits which can be prepared and measured accurately, and made to interact through high-quality quantum logic gates. 
Successful fault-tolerant operation requires certain minimum thresholds for the fidelity of state preparation, logic gates and state measurement. 
Qubit readout is vital, not only for the final output of the QC, but for the error-correction essential for its operation.  There is a trade-off 
between the error rate permitted and the number of extra qubits required to achieve fault-tolerant quantum error-correction (QEC), but typical studies 
require errors to be below $\ish 10^{-3}$ for realistic implementations~\cite{05:Knill,07:Steane}. A correction network typically uses more gates than 
individual qubit readouts, so readout error is less critical than gate error~\cite{07:Steane}. On the other hand, precise readout can be used to 
compensate for gate error. It is also crucial in a measurement-based QC~\cite{06:Nielsen}.

Trapped-ion QC approaches generally use qubits based either on a hyperfine transition within the ground atomic 
level~\cite{01:Monroe,05:Langer,06:Acton,07:Olmschenk,07:Lucas}, or on an optical transition between ground and metastable 
states~\cite{07:Wunderlich}. State preparation is implemented by optical pumping. State measurement is achieved by repeatedly exciting a cycling 
transition which involves one of the qubit states but not the other, and measuring whether or not the ion fluoresces~\cite{80:Wineland}. The accuracy 
with which the measurement can be performed depends on the rate at which fluorescence photons can be detected for the ``bright'' qubit state, compared 
with the rate at which the ``dark'' state gets pumped to the bright manifold (e.g.\ by off-resonant excitation), or vice versa. The fact that many 
fluorescence photons can typically be emitted by the ion before the undesired pumping process occurs allows for high-fidelity single-shot measurement 
even though the absolute photon collection efficiency can be poor (typically \ish 0.1\%). Measurement fidelities of 98\%--99\% have been 
reported~\cite{01:Monroe,06:Acton,07:Olmschenk,07:Wunderlich}. In recent work, an optical qubit was measured indirectly by repetitive measurement on 
an ancilla qubit held in the same ion trap, yielding a fidelity of 99.94\% in $\ish 12\ms$~\cite{07:Hume}.

\begin{figure}
\includegraphics[width=\onecolfig]{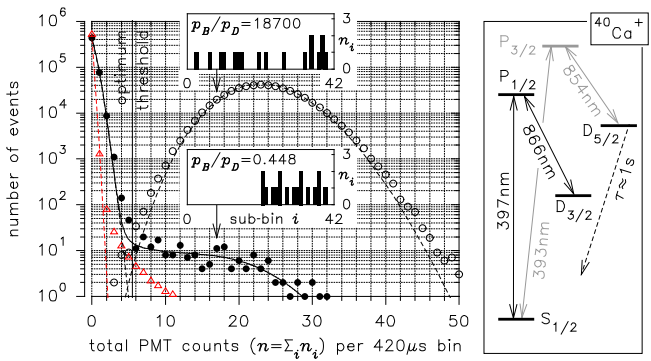}
\caption{%
\Ca{40} photon-count histograms with $t_b=420\us$, for bright \lev{S}{1/2} state ($\circ$) and dark \lev{D}{5/2} state ($\bullet$) preparations. 
Above-threshold events in the dark histogram give $\epsilon_D$; below-threshold events in the bright histogram give $\epsilon_B$. Insets show 
time-resolved counts for two individual 17-photon events, one from each histogram, with likelihood ratios $p_B/p_D$. A PMT dark count histogram is 
also shown (\smalltriangle), normalized to the same area. Dashed curves are Poisson distributions with the same means as the histograms. The solid 
curve is the expected \lev{D}{5/2} distribution taking into account spontaneous decay.
}
\label{F:Ca40hist}
\end{figure}

\begin{figure}
\includegraphics[width=0.99\onecolfig]{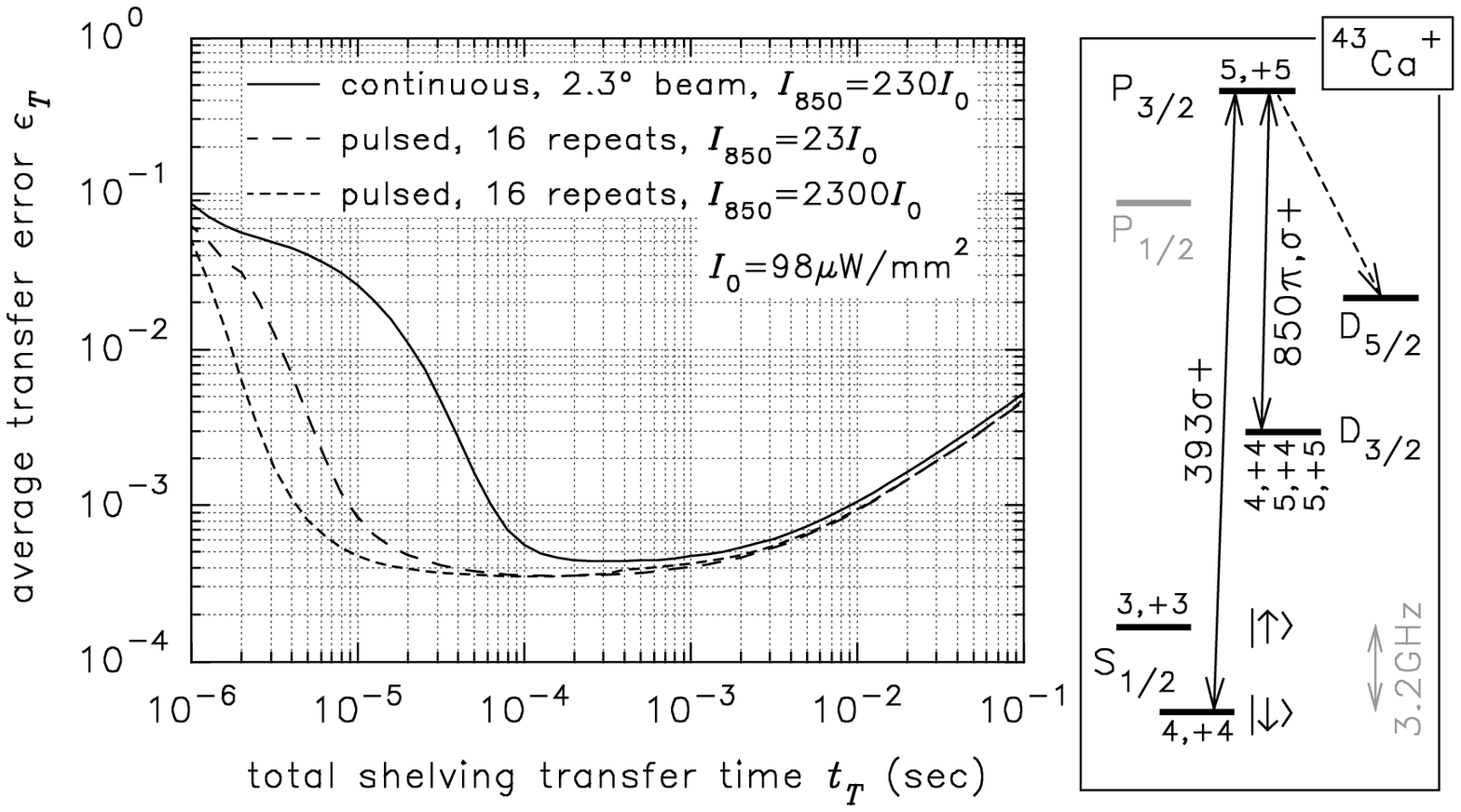} 
\caption{%
Rate equation simulations for the \Ca{43} shelving transfer $(\ket{\uparrow},\ket{\downarrow})\rightarrow(\ket{\uparrow},\lev{D}{5/2})$. The optimum 
average shelving transfer error $\epsilon_T$ is shown versus total time $t_T$ allowed for the transfer, for both continuous and pulsed methods. The 
393\nm\ intensity and (in the pulsed method) all pulse durations were numerically optimized for each $t_T$. The error increases at short $t_T$ due to 
higher 393\nm\ intensity causing off-resonant shelving of \ket{\uparrow}; at long $t_T$ the error increases because of decay from the \lev{D}{5/2} 
shelf. The minimum error is $3.5\times 10^{-4}$ at $t_T=160\us$; for $t_T=10\us$ an error of $4.8\times 10^{-4}$ is available. {\em Right:} Simplified 
\Ca{43} level diagram; states relevant to the shelving transfer are labelled by $F,M_F$.}
\label{F:Ca43shelving}
\end{figure}

In this Letter, we report {\em direct\/} high-fidelity measurement of firstly an optical qubit stored in the (\lev{4S}{1/2}, \lev{3D}{5/2}) 
levels~\footnote{The readout method is not sensitive to the choice of $M_J$.} of a single \Ca{40} ion  (fig.~\ref{F:Ca40hist}, right) and secondly a 
hyperfine qubit stored in the \lev{S}{1/2} $(F=3, F=4)$ sub-levels of a \Ca{43} ion (fig.~\ref{F:Ca43shelving}, right), where we first map the 
hyperfine qubit to the \Ca{43} optical qubit. Readout is achieved by driving the 
($\lev{S}{1/2}\leftrightarrow\lev{P}{1/2}\leftrightarrow\lev{D}{3/2}$) manifold and detecting the fluorescence from the 
$\lev{P}{1/2}\rightarrow\lev{S}{1/2}$ decay: fluorescence indicates the qubit was initially in the ``bright'' \lev{S}{1/2} state, absence of 
fluorescence that the qubit was in the ``dark'' metastable \lev{D}{5/2} state (lifetime $\tau=1168(7)\ms$~\cite{00:Barton}). We discuss state 
inference by simple photon-count thresholding and by maximum likelihood methods using time-resolved detection. The latter are applicable to a wide 
range of physical qubits~\cite{07:Gambetta}.

For the \Ca{40} measurements, a single ion is held in a Paul trap~\cite{00:Barton}, Doppler-cooled on the $\lev{S}{1/2}\leftrightarrow\lev{P}{1/2}$ 
transition by a 397\nm\ laser and repumped on $\lev{D}{3/2}\leftrightarrow\lev{P}{1/2}$ by an 866\nm\ laser. A photomultiplier (PMT) detects 
fluorescence from the ion with a net efficiency 0.19(2)\%. The ratio of ion fluorescence to background scattered laser light has a maximum value 
$\approx 690$ at low 397\nm\ intensity, but optimum readout fidelity is achieved at higher intensity because the increased fluorescence can be 
detected more rapidly compared with $\tau$. Mean photon count rates for fluorescence ($R_B$) and background ($R_D$) in this experiment were 
$R_B=55800\cps$ and $R_D=442\cps$ (including the PMT dark count rate of 8.2\cps\ at 20\degC).

The experimental sequence consists of preparing and measuring each qubit state in turn, and comparing the measurement outcome with the known 
preparation. To prepare the \lev{D}{5/2} ``shelf'' state a 393\nm\ beam is switched on to drive $\lev{S}{1/2}\leftrightarrow\lev{P}{3/2}$ for 1\ms; 
spontaneous decay populates \lev{D}{5/2} (the 866\nm\ beam is left on to empty \lev{D}{3/2}). The rate of transfer to the shelf was measured to be 
$[12(2)\us]^{-1}$. The 393\nm\ beam is extinguished and, 20(2)\us\ later, we start to collect PMT counts. Photons are counted for a bin time of 2\ms, 
which is divided into $200$ sub-bins of duration $t_s=10\us$. Two alternately-gated 10\MHz\ counters are used, so that there is negligible ($<50\ns$) 
dead-time between sub-bins. The counts $n_i$ from each sub-bin $i$ are available to the control computer \ish 5\us\ after the end of the sub-bin, 
should they be required for real-time processing, and are recorded in time order. Next, the ion is prepared in \lev{S}{1/2} by driving 
$\lev{D}{5/2}\leftrightarrow\lev{P}{3/2}$ for \ish 3\ms\ with an 854\nm\ laser beam, after which fluorescence is collected for a second 2\ms\ bin.

This sequence was repeated to give $2^{20}\approx 10^6$ trials, in half of which the ion was prepared in the dark \lev{D}{5/2} state and in half of 
which it was prepared in the bright \lev{S}{1/2} state. We define the average readout error to be $\epsilon=\frac{1}{2}(\epsilon_B+\epsilon_D)$, where 
$\epsilon_B$ is the fraction of experiments in which an ion prepared in the bright state was detected to be dark, and similarly for $\epsilon_D$. In 
the case of the dark state experiments, there are two small preparation errors arising from (i) the finite shelving rate of the 393\nm\ laser and (ii) 
the probability of spontaneous decay during the 20\us\ delay between state preparation and the start of the detection period. These give a 
contribution to $\epsilon_D$ of $(12\us+20\us)/\tau=0.28(2)\times 10^{-4}$. Known state preparation errors for the bright state are negligible, so the 
net contribution to $\epsilon$ is $0.14(1)\times 10^{-4}$. Values of $\epsilon$ given below are after subtraction of this quantity.

\begin{figure}
\includegraphics[width=\onecolfig]{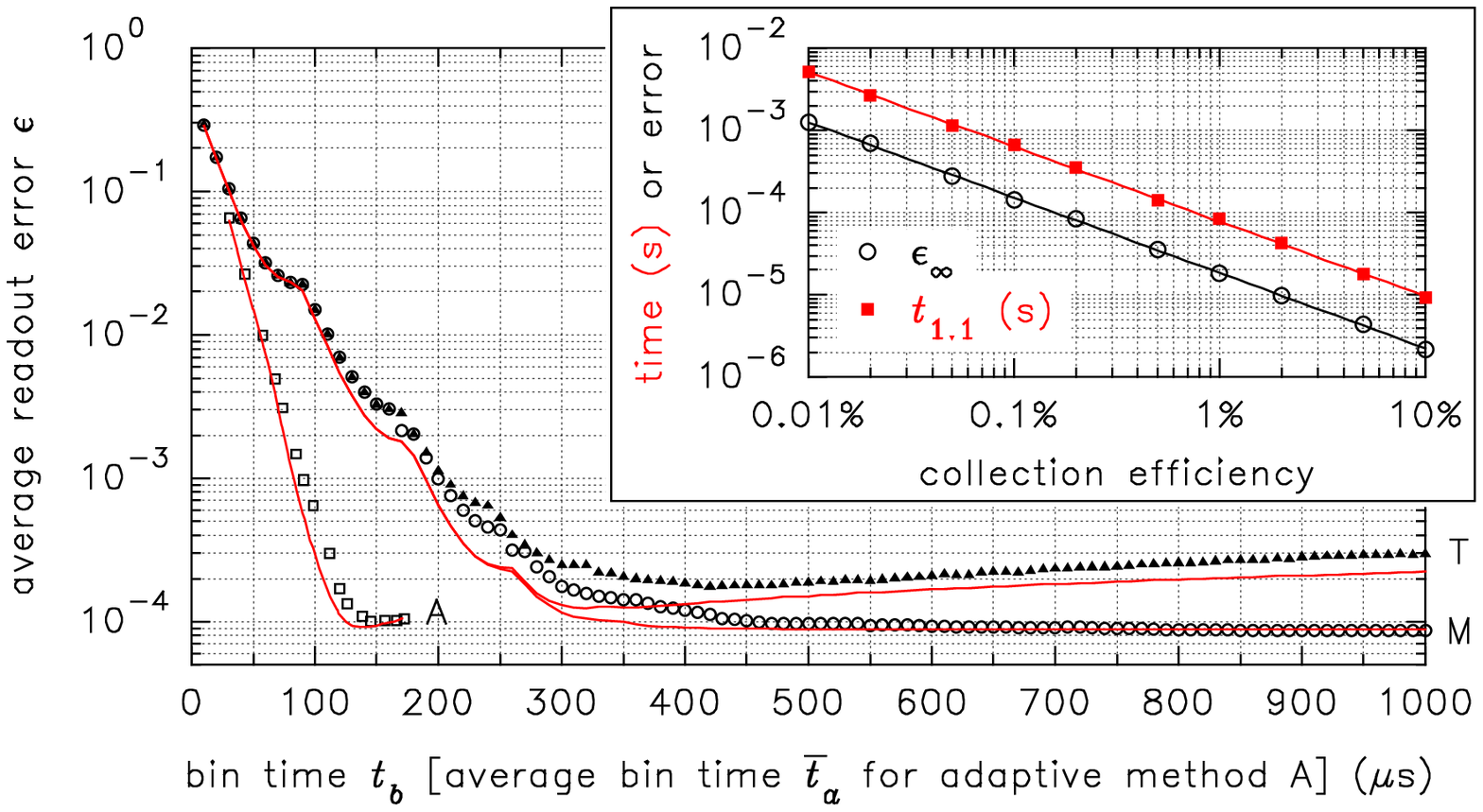}
\caption{%
Average readout error $\epsilon$ versus readout time for the \Ca{40} optical qubit, for three different analysis methods. Analyses of the experimental 
data are plotted as symbols; statistical uncertainty is at most the size of the symbols. The accompanying curves are simulations of $10^9$ trials 
using ideal Poisson statistics. Cusps arise from the discrete nature of photon counting. T: photon-count threshold method, where the threshold $n_c$ 
was optimized for each bin time. M: maximum likelihood method. A: adaptive maximum likelihood method. {\em Inset:} Simulations using method M, showing 
the asymptotic average error $\epsilon_\infty$ and the time $t_{1.1}$ required to reach $\epsilon=1.1\epsilon_\infty$ as functions of the net photon 
collection efficiency.
}
\label{F:errorVtime}
\end{figure}

Histograms of the number of photons $n=\sum_{i=1}^N n_i$ collected for the bright and dark state preparations are shown in fig.~\ref{F:Ca40hist}. Here 
the first $N=42$ sub-bins of the detection period were summed for a total detection time of $t_b=Nt_s=420\us$; this choice of $t_b$ and a threshold at 
$n_c=5\frac{1}{2}$ counts optimize the discrimination between the bright and dark histograms. The error $\epsilon$ as a function of $t_b$ is shown in 
fig.~\ref{F:errorVtime}, for $N=1\ldots 100$. As $t_b$ increases, $\epsilon$ first drops rapidly due to decreasing overlap between the two histograms, 
to a minimum $\epsilon=1.8(1)\times 10^{-4}$ at $t_b=420\us$, then rises slowly because of the increasing probability of decay from \lev{D}{5/2} 
during $t_b$.

At the optimum $t_b$, the majority of the error is due to the above-threshold events in the dark histogram: $\epsilon_D=3.2(3)\times 10^{-4}$. These 
mostly arise due to spontaneous decay during $t_b$, but we also observe above-threshold events from non-Poissonian PMT dark counts. A PMT dark count 
distribution is also shown in fig.~\ref{F:Ca40hist}: the highly non-Poissonian tail in the histogram probably arises as a result of cosmic 
rays~\cite{87:Teich}. The slow time-scale of luminescence in the PMT envelope excited by the cosmic rays makes it difficult to eliminate these events 
entirely. We estimate that, for the threshold method, cosmic ray events account for $\ish 20\%$ of $\epsilon_D$.

The threshold method does not make use of the arrival-time information of the photons. Using this information, we may hope to identify some of the 
events where decay from \lev{D}{5/2} occurs during $t_b$. For example, an ion decaying near the end of $t_b$ may give $>n_c$ counts but the fact that 
these occur in a burst at the end of the bin would suggest that this is more likely a decay event than a bright ion (compare fig.~\ref{F:Ca40hist} 
insets). The use of time-resolved measurement in the context of qubit readout was suggested in~\cite{01:Monroe} and modelled theoretically 
in~\cite{Th:Langer,07:Gambetta}; we follow a similar maximum likelihood treatment to Langer~\cite{Th:Langer}.

We calculate the likelihood $p_B$ that a given set of sub-bins $\{n_i\}$ could have been generated by a bright ion, and compare this with the 
likelihood $p_D$ that $\{n_i\}$ arose from an ion that was dark at the start of the detection period. If $p_B > p_D$ we infer that the ion was bright 
and vice versa. $p_B = P(\{n_i\}|\mbox{bright})$ is given simply by the product of probabilities $p_B = \Pi_{i=1}^N B(n_i)$ where $B(n_i)$ is the 
probability of observing $n_i$ counts from a bright ion (in the ideal case $B(n_i)$ is a Poisson distribution with mean count per sub-bin $R_B t_s$). 
The calculation of $p_D = P(\{n_i\}|\mbox{dark})$ is more involved, because we must sum over the possibilities that the ion decayed after or during 
the detection time $t_b=N t_s$. To a good approximation
\begin{equation}
p_D = \left(1-\frac{t_b}{\tau}\right) \prod_{i=1}^N D(n_i)
    + \left(\frac{t_s}{\tau}\right) \sum_{j=1}^N \prod_{i=1}^{j-1} D(n_i) \prod_{i=j}^N B(n_i)
\label{E:pD}
\end{equation}
where $(1-t_b/\tau)$ is the probability the ion decays after the last sub-bin, $(t_s/\tau)$ the probability it decays during any particular sub-bin 
$j$, and $D(n_i)$ the background count distribution with mean $R_D t_s$. Using a recursion relation
\begin{equation}
\begin{array}{ll}
\multicolumn{2}{l}{p_D = (1- t_b/\tau) M_N + (t_s/\tau) S_N}  \\  %
M_0 = 1, & M_k = M_{k-1} D(n_k)                               \\  %
S_0 = 0, & S_k = (S_{k-1} + M_{k-1}) B(n_k) %
\end{array}
\label{E:recurse}
\end{equation}
reduces the calculation of $p_D$ from O($N^2$) to O($N$) operations, making real-time readout much faster. In (\ref{E:pD}) we made the simplifying 
approximations that $t_s, t_b \ll \tau$, and that the mean count rate changes from $R_D$ to $R_B$ at the start of the sub-bin $j$ in which the ion 
decays; similar expressions are found if these approximations are not made, with negligible effect on the results.

The results of applying the maximum likelihood method to the same data set are also shown in fig.~\ref{F:errorVtime}, for $N=1\ldots 100$. For 
$B(n_i)$ and $D(n_i)$, Poisson distributions were convolved with the PMT dark count distribution~\footnote{We neglect time correlations caused by the 
cosmic rays.}. We see that for $t_b\gtish 200\us$ the maximum likelihood method gives a lower error than the photon-count thresholding method, tending 
to an asymptotic value $\epsilon=0.87(11)\times 10^{-4}$ (with $\epsilon_D=1.5(2)\times 10^{-4}$), or a fidelity 99.9913(11)\%. Furthermore, the 
maximum likelihood method does not require a precise choice of parameters ($n_c, t_b$) which must be determined from control data; we only need to 
know the independently measured rates $R_B$ and $R_D$, and to choose a sufficiently long $t_b$. The observed error is compared with that from a 
simulation of $10^9$ trials using ideal Poisson statistics in fig.~\ref{F:errorVtime}: although at short $t_b$ the experimental $\epsilon$ is greater 
due to super-Poissonian noise in the data, at long $t_b$ it converges to the simulation's asymptote at $\epsilon=0.89\times 10^{-4}$. We conclude 
that, at optimum $t_b$, the maximum likelihood method is less sensitive to experimental noise (e.g.\ cosmic ray events, drift in $R_B$) than the 
threshold method. 

We now discuss faster detection by using an adaptive version of the maximum likelihood technique~\cite{07:Hume}. In the preceding discussion, the 
detection bin length $t_b=N t_s$ was fixed and we evaluated the likelihoods $p_B, p_D$ at the end of the bin, deciding that the ion was dark if $p_D > 
p_B$. However, the absolute values of $p_B$ and $p_D$ also contain useful information. The {\em estimated\/} error probability $e_D$ that we have 
incorrectly deduced the ion to be dark when $p_D > p_B$ is given by Bayes' theorem~\cite{1763:Bayes}:
\[ e_D = 1 - P(\mbox{dark}|\{n_i\}) = 1 - \frac{P(\{n_i\}|\mbox{dark})}{P(\{n_i\})} = \frac{p_B}{p_B+p_D} \]
and similarly for $e_B$ in the case where $p_B\ge p_D$. By evaluating $p_D$ and $p_B$ at the end of each sub-bin, we can terminate the detection bin 
at $t_a < t_b$ when the estimated error probability $e_B$ or $e_D$ falls below some chosen cut-off threshold $e_c$. We also impose a cut-off time $t_c 
\le t_b$ in case the error threshold is not reached; $t_c$ is thus the worst case readout time. The result is that, for a given desired error level, 
the average detection bin length $\bar{t}_a$ is shorter than when the bin length is fixed. If $e_c$ is not reached, $e_B$ or $e_D$ still quantifies 
confidence in the measurement outcome, useful for QEC. Using the recursion relations (\ref{E:recurse}) we can evaluate $p_B, p_D$ in $<1\us$, while 
each sub-bin is in progress, so there is negligible time overhead.

For our conditions we find that, at little cost to $\epsilon$, faster detection can be obtained by omitting the effect of spontaneous decay in the 
analysis, i.e.\ replacing (\ref{E:pD}) by $p_D=\Pi_{i=1}^N D(n_i)$. An adaptive analysis of the same data set is shown in fig.~\ref{F:errorVtime}, for 
$t_c=500\us$ and a range of threshold values $10^{-6} \leq e_c \leq 10^{-1}$. For each $e_c$ we plot the experimentally measured average error 
$\epsilon$ against $\bar{t}_a$. For an average readout time of $\bar{t}_a=145\us$, the error reaches $\epsilon=1.0(1)\times 10^{-4}$, more than thirty 
times lower than the non-adaptive maximum likelihood method, or over three times as fast for the same $\epsilon$. The average readout time for the 
bright (dark) state is 72\us\ (219\us). 

For all of the analysis methods described, the bright \lev{S}{1/2} state can be detected more accurately than the dark \lev{D}{5/2} state because the 
known bright $\rightarrow$ dark transfer rate is so low (we measure it to be $<10^{-3}\persec$). The data imply the bright state can be detected with 
over $99.9998\%$ fidelity, $\epsilon_B < 2\times 10^{-6}$, in average time $\bar{t}_a=91\us$ (whilst retaining $\epsilon_D = 2.5(2)\times 10^{-4}$ in 
$\bar{t}_a=292\us$). This asymmetric readout error could be exploited for QEC.

The optical qubit offers high-fidelity readout but it suffers from two drawbacks: a finite lifetime $\tau$ and the need for high frequency stability 
in the optical domain. Qubits stored in hyperfine ground states avoid these problems and exhibit some of the longest coherence times ever 
measured~\cite{05:Langer,07:Lucas}. We turn now to the implementation of state detection for a qubit stored in the 
$\ket{\uparrow}=\hfslev{S}{1/2}{3,+3}$ and $\ket{\downarrow}=\hfslev{S}{1/2}{4,+4}$ hyperfine ground states of \Ca{43} (where the superscripts give 
the quantum numbers $F, M_F$). For readout, we first map this hyperfine qubit to the optical qubit by state-selective transfer 
$\ket{\downarrow}\rightarrow\lev{D}{5/2}$.

We Doppler-cool a single \Ca{43} ion on the $\hfslev{S}{1/2}{3}\leftrightarrow\hfslev{P}{1/2}{4}$ and 
$\hfslev{S}{1/2}{4}\leftrightarrow\hfslev{P}{1/2}{4}$ transitions. State preparation of \ket{\downarrow} is by optical pumping with 
$\sigma^+$-polarized light on the same transitions. Preparation of \hfslev{S}{1/2}{3,M_F} is achieved by extinguishing the 
$\hfslev{S}{1/2}{3}\leftrightarrow\hfslev{P}{1/2}{4}$ beam before the $\hfslev{S}{1/2}{4}\leftrightarrow\hfslev{P}{1/2}{4}$ beam: this prepares a 
statistical mixture of $M_F$ states rather than \ket{\uparrow} but the calculated effect on the measured readout error is $<10^{-4}$. To map 
$\ket{\downarrow}$ to the \lev{D}{5/2} shelf the ion is illuminated with 393\nm\ $\sigma^+$ light which drives 
$\ket{\downarrow}\leftrightarrow\hfslev{P}{3/2}{5,+5}$. This results in optical pumping to \lev{D}{5/2} or \lev{D}{3/2} with branching ratios 5.3\% 
and 0.63\% respectively~\cite{00:Barton}. Light at 850\nm\ containing $\sigma^+$ and $\pi$ polarizations is used to empty the relevant \lev{D}{3/2} 
states via the \hfslev{P}{3/2}{5,+5} state (fig.~\ref{F:Ca43shelving}, right). Assuming perfect repumping from \lev{D}{3/2}, the situation reduces to 
that for ions without low-lying D levels, but with the important difference that we only need to drive enough transitions to transfer \ket{\downarrow} 
to the shelf (rather than to collect fluorescence) before off-resonant excitation of \ket{\uparrow} occurs; since the 
$\lev{P}{3/2}\rightarrow\lev{D}{5/2}$ branching ratio is much greater than typical photon collection efficiencies this gives a significant advantage.

To optimize parameters the shelving transfer process was modelled by rate equations applied to the entire 144-state (4S, 4P, 3D) manifold. The optimum 
shelving method would be a repeated sequence of three laser pulses (393\nm\ $\sigma^+$, 850\nm\ $\sigma^+$, 850\nm\ $\pi$), since then there is 
negligible probability of the ion decaying to $\ket{\uparrow}$. The main limitation is off-resonant (by 3.1\GHz) excitation of 
$\ket{\uparrow}\leftrightarrow\hfslev{P}{3/2}{4,+4}$. Continuous excitation allows similar fidelity if the 850\nm\ $\pi$ component is weak (though 
with slower shelving; see fig.~\ref{F:Ca43shelving}). Accordingly, in the experiment, we used a single circularly-polarized 850\nm\ beam travelling at 
a small angle (2.3\degree) to the quantization axis, giving polarization components with intensities 
$(I_{\sigma^+},I_\pi,I_{\sigma^-})=(0.9992,0.0008,2\times 10^{-7})\times 230(70)I_0$, where the saturation intensity $I_0=98\unit{\uW/mm$^2$}$. The 
shelving transfer was accomplished with a single simultaneous $393\nm+850\nm$ pulse, with duration set to the optimum value predicted by the model, 
$t_T=400\us$. No improvement was found by varying $t_T$ or by using alternating 393\nm\ and 850\nm\ pulses (to avoid two-photon effects, which are not 
modelled by the rate equations). After shelving, state detection proceeds as for \Ca{40}; photon-count thresholding with $t_b=2\ms$ was used as the 
optical qubit readout is not the limiting factor.

The net average readout error for the hyperfine qubit was measured from 20000 trials to be 
$\epsilon\sub{hfs}=\frac{1}{2}(\epsilon_\uparrow+\epsilon_\downarrow)=2.3(3)\times 10^{-3}$, a fidelity of 99.77(3)\%. (The error for the 
\ket{\uparrow} state was $\epsilon_\uparrow=1.7(4)\times 10^{-3}$.) A separate experiment measured the optical qubit readout error to be $1.0(2)\times 
10^{-3}$, significantly poorer than for \Ca{40} due to a lower fluorescence rate $R_B=7500\cps$ (caused partly by coherent population trapping 
effects, which could be eliminated by polarization modulation techniques~\cite{00:Boshier}). This implies an average error $\epsilon_T=1.3(4)\times 
10^{-3}$ for the shelving transfer. This is somewhat above the modelled value of $0.44\times 10^{-3}$; possible reasons for this include imperfect 
circular polarization of the 850\nm\ beam, imperfect population preparation, and broadening of the 393\nm\ transition (e.g.\ due to finite laser 
linewidth).

In conclusion, we demonstrate fast, direct, single-shot readout from optical and hyperfine trapped-ion qubits at fidelities comparable with those 
required for a fault-tolerant QC. An adaptive detection method reduced the optical qubit readout error by a factor $\approx 35$ for a given average 
detection time, reaching 99.99\%\ fidelity in 145\us, with negligible time overhead due to the classical control system. A simple and robust method 
for efficient readout of hyperfine qubits, capable of comparable fidelity/time performance, was proposed and implemented. An increase in the photon 
collection efficiency by an order of magnitude, to 2\%, would speed up both readout methods, and reduce the optical qubit readout error, by a similar 
factor (fig.~\ref{F:errorVtime}, inset).

We thank C.~Langer and J.~P.~Home for helpful discussions. This work was supported by EPSRC (QIP IRC), DTO (contract W911NF-05-1-0297), the European 
Commission (``SCALA'', ``MicroTrap'') and the Royal Society.

\end{document}